# LEGAL DOCUMENT RETRIEVAL ACROSS LANGUAGES: TOPIC HIERARCHIES BASED ON SYNSETS

Horizon 2020 No 780247 - TheyBuyForYou

**Authors:**

Badenes-Olmedo, Carlos (Universidad Politécnica de Madrid)

Redondo-Garcia, Jose Luis (Amazon Research)

Corcho, Oscar (Universidad Politécnica de Madrid)

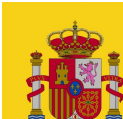
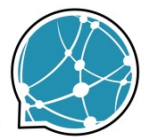

# LEGAL DOCUMENT RETRIEVAL ACROSS LANGUAGES:

# TOPIC HIERARCHIES BASED ON SYNSETS[*]

Cross-lingual annotations of legislative texts enable us to explore major themes covered in multi-lingual legal data and are a key facilitator of semantic similarity when searching for similar documents. Multilingual probabilistic topic models have recently emerged as a group of semi-supervised machine learning models that can be used to perform thematic explorations on collections of texts in multiple languages. However, these approaches require theme-aligned training data to create a language-independent space, which limits the amount of scenarios where this technique can be used. In this work, we provide an unsupervised document similarity algorithm based on hierarchies of multi-lingual concepts to describe topics across languages. The algorithm does not require parallel or comparable corpora, or any other type of translation resource. Experiments performed on the *English*, *Spanish*, *French* and *Portuguese* editions of JCR-Acquis corpora reveal promising results on classifying and sorting documents by similar content.

## 1. INTRODUCTION

Searching for similar documents among legal data accross Europe requires multi-language techniques to be performed. For many languages we may not have access to translation dictionaries or a full translation system, or they can be expensive to apply in an online search system. In such situations it is useful to rely on smaller annotation units derived from the text so the full content doesn't need to be translated, for instance by finding correspondences with regard to the topics discussed. In this case, it may be advisable to automatically learn cross-lingual topics to browse multi-lingual document collections.

Within the European project *TheyBuyForYou[1]*, we have worked on an unsupervised way of building cross-lingual labels from sets of cognitive synonyms (synsets) to establish relations between language-specific topics. This avoids the need for *parallel* corpus (i.e sentence-aligned documents) or *comparable* corpus (i.e theme-aligned documents), that is a requirement of the recently emerged multilingual probabilistic topic models (MuPTM) [1]. The cross-lingual labels can be used for large-scale document retrieval tasks in multi-lingual corpora.

---

[*] *This paper is an extension of our previous work [3] by adding Portuguese texts into the evaluation.*

[1] *https://theybuyforyou.eu*

## 2. APPROACH

Documents are represented as data points in a low-dimensional latent space created by probabilistic topic models for each language separately (*figure 1*). Topics are annotated with cross-lingual labels created from a list of synset [2] retrieved from the Open Multilingual WordNet[2]. Word by word are queried in WordNet to retrieve its synsets. The final set of synsets for a topic is the union of the synsets from the individual top-words of a topic (*top5* based on empirical evidences) [3].

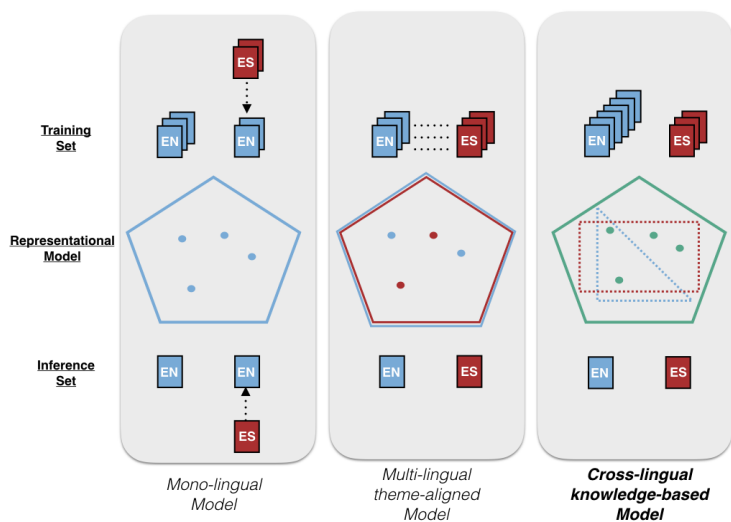

*Figure 1 Mono-lingual, multi-lingual and cross-lingual topic-based approaches*

Since exact similarity computations are unaffordable for neighbours detection tasks ($O(n^2)$) [7], a hashing algorithm to group similar documents and to preserve the notion of topics has been proposed [4]. It defines a hierarchical set-type data where each level of the hierarchy indicates the importance of the topic according to its distribution (*figure 2*). The knowledge provided by the topics to describe the documents is maintained and an efficient exploration of document collections on a large scale can be performed. Documents are annotated with the relation between topics inside each hierarchy level. In this feature space, the proposed distance metric is based on the *Jaccard* coefficient which computes the similarity of sets by looking at the relative size of their intersection. More specifically, the similarity metric used to compare the hash codes created from set of topics is the sum of the Jaccard distances for each hierarchy level, i.e. for each set of topics [4].

---

[2] *http://compling.hss.ntu.edu.sg/omw/*

## 3. Evaluation

The JRC-Acquis[3] dataset is used to compare the performance of this unsupervised algorithm with a semi-supervised algorithm based on MuPTM. It is a collection of legislative texts written in 23 languages that have been manually classified into subject domains according to the EUROVOC[4] thesaurus. The English, Spanish, French and Portuguese editions (about 20,000 documents per edition) of the corpora were used for each language-specific model. The EUROVOC taxonomy was pre-processed to satisfy the *topic independence* assumption of LDA [5] models, by using hierarchical relations. The initial 7,193 concepts

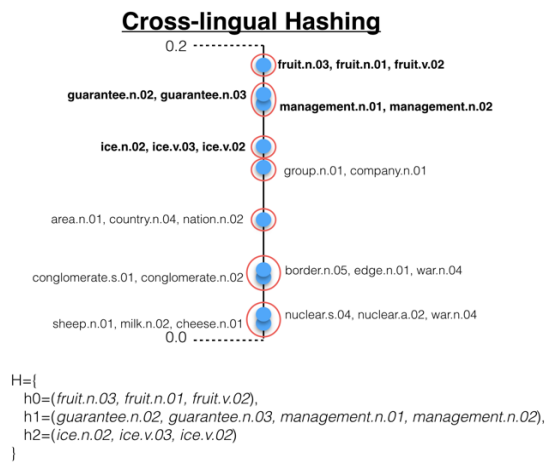

*Figure 2 hash-expression (H) of a document based on WordNet-synset annotations created from the top words of each topic distribution*

from 21 domain areas such as politics, law or economics were reduced to 452 categories, that are independent and can be used to train the topic models. Finally, in addition to LDA models[5], LabeledLDA [6] models[6] were also created to force the correspondence between those categories and the latent topics of the model.

Documents were pre-processed (Part-of-Speech filtering and lemmatized format) by the librAIry[7] NLP service to build a suitable dataset for the models. Theme-aligned probabilistic topic models in English, Spanish, French and Portuguese share the topics but not their definitions (i.e words) (see *table 1*).

| Topic3@EN<br>*Communication Systems* | Topic3@ES<br>*Sistema de Comunicación* | Topic26@FR<br>*Systeme de Comunicacion* | Topic10@PT<br>*Communication Systems* |
|---|---|---|---|
| radio | equipo | communications | rede |
| equipment | red | reseaux | comunicação |
| network | comunicación | electroniques | electrónico |
| communication | espectro | acces | acesso |
| regulatory | electromagnético | telecommunications | utilizador |

*Table 1 Theme-aligned topics described by top 5 words based on EUROVOC annotations*

---

[3] *https://ec.europa.eu/jrc/en/language-technologies/jrc-acquis*

[4] *http://eurovoc.europa.eu/*

[5] *http://librairy.linkeddata.es/jrc-{en|es|fr|pt}-model-unsupervised*

[6] *http://librairy.linkeddata.es/jrc-{es|en|fr|pt}-model*

[7] *http://librairy.linkeddata.es/nlp*

### 3.1 Cross-lingual Document Retrieval Results

Given a set of documents and a query document, the task is to filter and rank the documents according to their relevance (i.e. semantic similarity) to the query text regardless of the language used. A ground-truth set grouping the documents that share the same EUROVOC codes is considered from the query document. A collection of 1,000 randomly selected documents (monolingual, bi-lingual and multi-lingual) are annotated by the category-based (semi-supervised model) and synset-based (unsupervised model) algorithms. We evaluate the performance of the algorithms in terms of precision@3, precision@5 and precision@10 (*tables 2 and 3*)

|       | en       |        | es       |        | fr       |        | pt       |        |
|-------|----------|--------|----------|--------|----------|--------|----------|--------|
|       | category | synset | category | synset | category | synset | category | synset |
| p@3   | **0.84** | 0.83   | **0.81** | 0.78   | **0.83** | 0.74   | **0.79** | 0.78   |
| p@5   | **0.82** | 0.80   | **0.79** | 0.75   | **0.80** | 0.72   | **0.77** | 0.75   |
| p@10  | **0.77** | 0.76   | **0.75** | 0.73   | **0.77** | 0.58   | **0.72** | 0.71   |

*Table 2: Information retrieval performance of the categories-based and synset-based topic alignment algorithms in monolingual document collections.*

|       | es-en    |        | es-pt    |        | pt-en    |        | en-es-fr-pt |        |
|-------|----------|--------|----------|--------|----------|--------|-------------|--------|
|       | category | synset | category | synset | category | synset | category    | synset |
| p@3   | **0.84** | 0.79   | **0.80** | 0.78   | **0.82** | 0.81   | **0.81**    | 0.69   |
| p@5   | **0.82** | 0.76   | **0.77** | 0.76   | **0.80** | 0.78   | **0.78**    | 0.67   |
| p@10  | **0.78** | 0.73   | **0.75** | 0.72   | **0.77** | 0.75   | **0.73**    | 0.62   |

*Table 3: Information retrieval performance of the categories-based and synset-based topic alignment algorithms in multi-lingual document collections.*

Although the precision values are lower than those obtained by semi-supervised approximation, they are sufficiently promising (around 0.75) to think that introducing improvements in the natural language processing would increase the quality of the WordNet-synset annotations derived from the most representative words of each topic (precision values close to 0.8 in the English corpus).

## 4. Conclusion

The algorithm has proved to perform close to the semi-supervised algorithm in information retrieval tasks, which makes us think that the process of topic annotation by set of synonyms can be improved to filter those elements that are not sufficiently representative. Our future lines of work will go in that direction, incorporating context information to identify the most representative synset for each topic.

## 5. Acknowledgments

This research was supported by the European Union's Horizon 2020 research and innovation programme under grant agreement No 780247: TheyBuyForYou.